\documentclass[fleqn,usenatbib]{mnras} 

\usepackage[T1]{fontenc}

\DeclareRobustCommand{\VAN}[3]{#2}
\let\VANthebibliography\thebibliography
\def\thebibliography{\DeclareRobustCommand{\VAN}[3]{##3}\VANthebibliography}

\usepackage{graphicx}	% Including figure files
\usepackage{amsmath}	% Advanced maths commands
\usepackage{amssymb}	% Extra maths symbols
\usepackage{bm}

\newcommand{\pppt}{$\mathrm{P^3T}$}
\newcommand{\eesc}{e_{\mathrm{kin,esc}}}
\newcommand{\rin}{r_{\mathrm{in}}}
\newcommand{\rout}{r_{\mathrm{out}}}
\newcommand{\rij}{r_{ij}}
\newcommand{\ma}{m_{\mathrm{1}}}
\newcommand{\mb}{m_{\mathrm{2}}}
\newcommand{\mave}{\langle m \rangle}
\newcommand{\ab}{a_{\mathrm{m}}}
\newcommand{\rg}{r_{\mathrm{g}}}
\newcommand{\eerr}{\langle \mathcal{E} \rangle}
\newcommand{\dtl}{\Delta t_{\mathrm{l}}}
\newcommand{\dtsm}{\Delta t_{\mathrm{s,max}}}
\newcommand{\dtsi}{\Delta t_{\mathrm{s,i}}}
\newcommand{\dts}{\Delta t_{\mathrm{s}}}
\newcommand{\trh}{T_{\mathrm{rh}}}
\newcommand{\tcr}{T_{\mathrm{cr}}}
\newcommand{\rh}{R_{\mathrm{h}}}
\newcommand{\rc}{R_{\mathrm{c}}}

\newcommand{\nesc}{N_{\mathrm{esc}}}

\title[The reliability of simulations]{On the reliability of simulations of collisional stellar systems}

\author[Long Wang et al.]{
Long Wang,$^{1,2}$\thanks{E-mail:long.wang@astron.s.u-tokyo.ac.jp},
David M. Hernandez $^{3}$
\\
$^{1}$Department of Astronomy, School of Science, The University of Tokyo, 7-3-1 Hongo, Bunkyo-ku, Tokyo, 113-0033, Japan \\
$^{2}$RIKEN Center for Computational Science, 7-1-26 Minatojima-minami-machi, Chuo-ku, Kobe, Hyogo 650-0047, Japan\\
$^{3}${Harvard--Smithsonian Center for Astrophysics, 60 Garden St., MS 51, Cambridge, MA 02138, USA} \\
}

\date{Accepted XXX. Received YYY; in original form ZZZ}

\pubyear{2015}

\begin{document}
\label{firstpage}
\pagerange{\pageref{firstpage}--\pageref{lastpage}}
\maketitle

\begin{abstract}
%\change{I enjoyed the paper!  I will write comments in red.  Please check my edits using the History function in Overleaf.  Feel free to reject or change my edits.  I decided not to edit your English grammar in this draft, maybe in the next draft.  But please try to correct the grammar yourself first, there are many minor problems.  Also, I wanted to include some more references, you don't want to use a .bib file?  It seems inputting bibliography by hand is a bit tedious.}
  It is well known that numerical errors grow exponentially in $N$-body simulations of gravitational bound stellar systems, but it is not well understood how the accuracy parameters of algorithms affect the physical evolution in simulations.
  By using the hybrid $N$-body code, \textsc{petar}, we investigate how escapers and the structure evolution of collisional stellar systems (e.g., star clusters) depend on the accuracy of long-range and short-range interactions.
   We study a group of simulations of ideal low-mass star clusters in which the accuracy parameters  are varied.  
   We find that although the number of escapers is different in individual simulations, its distribution from all simulations can be described by Poisson statistics.
  The density profile also has a similar feature.
  %\change{This sounds strange, that same initial conditions can give different behavior, it would be good to say the parameters change slightly}, 
  By using a self-consistent set-up of the accuracy parameters for long- and short-range interactions, such that orbits are resolved well enough, the physical evolution of the models is identical.
%  To ensure the accuracy of short-range interaction is much more important than that of long-range interactions, as by reducing the former can result in completely .
  %\change{Is it possible to do low accuracy short-range but self-consistent?}.
  But when the short-range accuracy is too low, a nonphysical dynamical evolution can easily occur; this is not the case for long-range interactions. 
  This strengthens the need to include a proper algorithm (e.g. regularization methods) in the realistic modelling of collisional stellar systems.
  We also demonstrate that energy-conservation is not a good indicator to monitor the quality of the simulations. 
  The energy error of the system is controlled by the hardest binary, and thus, it may not reflect the ensemble error of the global system.
  %\change{(The reference cannot appear in abstract in MNRAS.) 
  %This agrees with results of \citeauthor{Hernandez2020} in that improved energy errors did not lead to improved ensemble errors.
%  }
\end{abstract}


%%%%%%%%%%%%%%%%%%%%%%%%%%%%%%%%%%%%%%%%%%%%%%%%%%

%%%%%%%%%%%%%%%%% BODY OF PAPER %%%%%%%%%%%%%%%%%%

\section{Introduction}
% multi timescale issue
The star-by-star numerical $N$-body simulations of collisional stellar systems, such as star clusters, are difficult due to the existence of compact few-body systems.
``Collisional'' means that close encounters between stars significantly affect the dynamical evolution of the system \citep[e.g.,][]{Binney1987}.
In a smooth potential like the Galactic potential, a star has a certain orbit around the Galaxy; it will maintain the same orbit unless the potential evolves.  This is a collisionless system.
In a collisional system, close encounters perturb the motions of stars; after a while, the stars will depart from their original orbits.
Such a process is referred as relaxation.
Relaxation drives energy exchange among stars.
If we use thermal dynamics to describe the evolution of gravitational bound systems, we can find that the heat capacity of the system is negative.
As the energy transfers from the core to the halo of the system, the core contracts while the temperature gradients along the radial direction of the system increase. 
This subsequently accelerates the contracting process and drives the core collapse.
%Then, the core collapse occurs driven by this gravothermal instability.
%Without any  \change{is feedback the right word?}, all stars in the center will merge to single object.
{However, when a binary exists in the core, it can encounter stars with separation comparable to the binary apo-center distance.  Then, it can transfer its binding energy as a heating source. }
%Finally, stars in the core will not merge to one object because of the binary heating.
This binary heating prevents an infinite collapse of the core.
%The Heggie-Hills law \citep{Heggie1975,Hills1975} suggests that in a stellar system, tight binaries tend to be tighter after interacting with others.
%Such interactions can transfer the binding energy of binaries out to prevent infinite shrinking of the core.
Therefore, binaries play a key role to control the long-term evolution of the system.
A realistic numerical simulation of a collisional stellar system must accurately treat the dynamical evolution of binaries.

However, as tight binaries shrink after interactions, their periods can become as short as days, which must be resolved.
Since the whole system evolves in a Gyr timescale, the simulation can become very time-consuming.
To solve this issue, $N$-body methods need to introduce approximations to avoid expensive computing.
For example, the state-of-the-art code, \textsc{nbody6} \citep{Aarseth2003}, ignores the long-range tidal force to binaries.
Thus, if a binary has no neighbor, it is treated as an isolated binary with a pure Kepler orbit.
Since the phase information of the binary is not important for the dynamical evolution of the system, \textsc{nbody6} does not evolve the isolated binary until a neighbor comes.
This can significantly reduce the computing cost. 

In a recently developed $N$-body code, \textsc{petar}, the slow-down method \citep{Mikkola1996,Wang2020b} and the tidal-tensor method \citep{Wang2020c} are used together.
Since the cumulative tidal effect of the whole system on a binary can be important, \textsc{petar} does not include isolated binaries like \textsc{nbody6}.
Instead, artificial particles are created around each tight binary to measure the local tidal tensor in a certain time interval.
Then, the tensor is used to evaluate the long-range tidal force to the binary.
Meanwhile, short-range forces from neighbors are directly calculated.
Since there are no isolated binaries, the orbit of each short-period binary needs to be integrated.
The slow-down method is implemented to avoid expensive computing.
Depending on the ratio between the perturbation and the internal force of a binary, a slow-down factor ($k>1$) is estimated. 
Then, the internal motion of the binary is reduced by $k$ times, while the perturbation force is increased by the same factor.
It can be mathematically proved that the slow-down method ensures a correct secular motion with a cost of losing the phase information.
In the example of a hyperbolic encounter in \cite{Wang2020b}, $k$ can be as large as $3\times10^4$.
Thus, the total integration steps are reduced by a few orders of magnitude compared to that of a conventional integrator.
%However, the slow-down Hamiltonian has a different definition of energy, thus the physical energy conservation is broken for this method \citep{Wang2020b}.

These methods cannot solve the problem of the singularity due to the of inverse square law of the Newtonian force.
To follow the motion of an eccentric binary, a traditional integrator, such as the fourth-order Hermite method, needs to use an extremely small time step at the peri-center in order to obtain an enough accuracy.
This not only is computationally expensive, but also leads to a large numerical error after many orbits. 
Such an error can completely change the orbit of the binary and even unbind it.
To solve this issue, a group of algorithms, so called ``regularization methods'', were introduced. 
The famous ones are the KS regularization \citep{KS1965}, the Burdet-Heggie regularization \citep{Burdet1967,Burdet1968} and the time-transformed symplectic leapfrog integrator \citep[or algorithmic regularization;][]{Preto1999,Mikkola1999}.
The coordinate transformations are used in these algorithms to remove the singularity. 
Thus, the efficiency and accuracy are significantly improved.
However, coupling such algorithms into $N$-body codes requires introducing a criterion to select the particles for which the regularization applies.
The force from nearby particles to the regularized group are treated as an external perturbation.
The criterion needs to be well determined so that the perturbation is much smaller than the internal forces of the binary.
This is important for the KS regularization method, which assumes a weak perturbation.
Meanwhile, the start and termination of the regularization introduces additional numerical errors.
Thus, the criterion also needs to ensure the switching is not frequent.
Determining a suitable criterion to ensure both computing efficiency and accuracy of integration for all kinds of conditions is extremely challenging.
Neither \textsc{nbody6} nor \textsc{petar} has a perfect solution.
%It is very difficult to ensure a high degree of energy conservation in such the hybrid codes.

There are only a few studies that have investigated how the numerical errors affect the physical properties in the $N$-body simulations of collisional stellar systems.
\cite{Heggie1991} studied how the formation of binaries, the evolution of half-mass radius and the number of escapers depend on the accuracy of the integrator.  They used a force polynomial combined with individual time steps \citep[\textsc{nbody1};][]{Aarseth2003}.
They found that, in an $N$-body model, when changing the parameter ($\eta$) to determine the time step, the formation time of a binary, and the evolution of the half-mass radius remain consistent.  However, a noticeable difference in the number of escapers is seen.
Since \textsc{nbody1} uses a softening potential to avoid singularities, close encounters were not properly treated.
This might affect the number of escapers. 
\cite{PZ2014} investigates how the accuracy of integration affects the few-body motions (the decay time) and found that small errors in individual simulations can finish at completely different orbits due to chaos, but the statistical mean of many repeating simulations can converge to the correct result. 
\cite{Hernandez2020} investigated the long-term evolution of planetary systems. 
They found that the statistics of action variables are accurate as long as all orbits are properly resolved; e.g., $~16$ steps per effective pericenter period.  Furthermore, \cite{Hernandez2021} finds that ensemble averaging improves measurement of action-like planetary quantities.

Energy conservation is frequently used to indicate the quality of the $N$-body simulations.
However, it is not a guarantee that good energy conservation means a correct physical evolution \citep[e.g.][]{Heggie1991}.
In collisional stellar systems, one tight binary can contain most of the energy in the entire system.
A small energy error from the binary can override any systematic error in the system.
Thus, an energy conservation check can be deceptive.

In this work, by performing a large amount of simulations for a low-mass star clusters using \textsc{petar} code,
we investigate how the statistical properties of simulations, especially the escapers, depends on the accuracy of long-range and short-range interactions.
{The results show that, although individual models in the groups have significant differences, the distributions of the models converge.}
In Section~\ref{sec:method}, we introduce the numerical algorithms used in \textsc{petar}.
In Section~\ref{sec:init}, we present the initial conditions of the models.
Then, a comparison for the evolution of the half-mass radius, the core radius, the properties of escapers, and the energy errors are presented in Section~\ref{sec:result}.
Finally, we make conclude in Section~\ref{sec:conclusion}.
%There are two important timescales to describe the dynamical status of such systems, crossing time ($\tcr$) and relaxation time ($\trh$).
%$\tcr$ represents how fast a star cross the entire system (or orbital period).
%$\trh$ indicates the time when a star lose the memory of its original orbit after interacting with surrounding stars.
%A star near the half-mass radius of a star cluster has a orbital period in the order of Myr, but a tight binary can have a period of days.
% necessary of regularization algorithm in stellar dynamics

\section{Method}
\label{sec:method}

In this work, we use \textsc{petar} to carry out numerical simulations \citep{Wang2020c}.
\textsc{petar} is a hybrid $N$-body code that combines three integration methods:
\begin{itemize}
    \item The Barnes-Hut tree \citep{Barnes1986} is used to calculate long-range forces between particles, which are integrated with a second-order symplectic leap-frog integrator (hereafter referred as ``PT'').
    \item The fourth-order Hermite integrator with block time steps \citep[e.g.,][hereafter referred as ``PP'']{Aarseth2003} is applied to integrate the orbits of stars and the centers-of-mass of multiple systems with short-range forces.
    \item The slow-down algorithmic regularization method \citep[SDAR;][]{Wang2020b} is used to integrate the multiple systems, such as hyperbolic encounters, binaries and hierarchical few-body systems.
\end{itemize}

When there are no binaries in the stellar system, the long-range forces are the most expensive computing part in simulations.
The direct pair-force summation for all particles in the PP method requires a computing cost of $O(N^2)$, while the cost for the PT method scales as $O(N\log{N})$.
However, using pure PT with a second-order leap-frog integrator to accurately handle the short-range interactions requires a small shared time step.
This can be even more expensive than the PP method for simulating the collisional stellar systems.
A hybrid algorithm, \pppt, combines the PT and the PP methods to include both their benefits of efficiency and accuracy.
This is done via the Hamiltonian splitting for a system with $N$ particles \citep[e.g.,][]{Oshino2011}:
\begin{equation}
  \begin{aligned}
  H_{\mathrm{s}} =& \sum_{i=1}^N\frac{p_i^2}{2 m_i} - \sum_{i<j}^N \frac{G m_i m_j}{\rij} W(\rij) \\
  H_{\mathrm{l}} =& \sum_{i=1,i<j}^N  \frac{G m_i m_j}{\rij} [1-W(\rij)] .\\
  \end{aligned}
  \label{eq:hsplit}
\end{equation}
where $H_{\mathrm{s}}$ includes the short-range interactions and kinetic energies and $H_{\mathrm{l}}$ indicates the long-range interaction; $p_i$ and $m_i$ are, respectively, the momenta and mass of particle $i$; $\rij$ is the separation between the $i$ and $j$ particles; $G$ is the gravitational constant; $W(\rij)$ is a changeover function to smoothly transfer pieces between $H_{\mathrm{s}}$ and $H_{\mathrm{l}}$.

\textsc{petar} uses a mass-dependent eighth-order polynomial changeover function for each particle.  
The order of this polynomial can have a significant effect on the accuracy of chaotic solutions \citep{Hernandez2019a,Hernandez2019b}.
In this work, we only investigate equal-mass star clusters, thus the changeover function is the same for all particles during one simulation.
The simplified changeover function for the equal-mass condition can be described as
\begin{equation}
  W(x)  = 
  \begin{cases}
    \beta (1-2 x) & (x\le 0) \\
    \beta (1-2 x) - 1 + f(x) & (0 < x < 1)\\
    0  & (x\ge 1) \\
  \end{cases}
\end{equation}
where
\begin{equation}
  \begin{aligned}
    f(x) & = 1 + \beta x^5 \left( 14 - 28 x + 20 x^2 -5 x^3 \right) \\
    x & = \frac{\rij-\rin}{\rout-\rin} \\
    \beta & = \frac{\rout-\rin}{\rout+\rin}. \\  
  \end{aligned} 
\end{equation}
$\rin$ and $\rout$ are the inner and outer boundaries of the changeover function, respectively.
These are determined at the beginning of the simulation.
We fix $\rout/\rin$ to be $10$ in our simulations, thus we only specify $\rout$.

The PP method uses an individual time step for each particle.
It is calculated by \citep{Aarseth2003,Oshino2011},
\begin{equation}
    \dtsi = \mathrm{min}\left ( \eta \sqrt{\tfrac{\sqrt{\left|\bm{A}^{(0)}_i\right|^2 + A^2_{0}} \left|\bm{A}^{(2)}_i\right| + \left|\bm{A}^{(1)}_i\right|^2}{\left|\bm{A}^{(0)}_i\right| \left|\bm{A}^{(3)}_i\right| + \left|\bm{A}^{(2)}_i\right|^2} }, \dtsm \right),\\
  \label{eq:dth}
\end{equation}
where $\bm{A}^{(0)}_i$ is the acceleration of a particle ($i$), $\bm{A}^{(j)}_i$ is its $j$-order time derivative, $A_{0}$ is a smoothing parameter with the unit of acceleration.  It avoids an unnecessarily small step when all the other terms are close to zero. $\dtsm$ is the maximum time step.
In our simulations, {$\eta=0.1$, a choice used in previous works} \citep[e.g.,][]{Aarseth2003}, and $A_{0}=0.1\mave/\rout^2$, where $\mave$ is the local average stellar mass.

To switch on the SDAR method, a mass-dependent radial criterion, $\rg$, is used to select the members of multiple systems.
In this work $\rg$ is uniform for all particles and $\rg < \rout$.

To deal with large $N$-body simulations, \textsc{petar} is developed by using the parallelization framework for developing particle simulation codes \citep[\textsc{fdps};][]{Iwasawa2016,Iwasawa2020}.
\textsc{fdps} provides the parallel-computing support for the PT part by using MPI and OpenMP software.
\textsc{petar} also supports the use of special accelerators, such as GPUs (using the CUDA language), SIMD (AVX, AVX2, AVX512) and Fujitsu A64FX (in the Fugaku supercomputer), to speed up the force calculation.

The hybrid integration method of \textsc{petar} is very suitable for this research because we can distinguish the impacts of long-range relaxation and short-range close encounters on the evolution of collisional stellar systems.

\subsection{Initial conditions}
\label{sec:init} 

In this work, we use the same initial conditions of a star cluster for all simulations.
The cluster contains $1000$ equal-mass particles.
The positions and the velocities of particles are randomly sampled from the Plummer distribution \citep[][]{Plummer1911}.
Initially, the system is in virial equilibrium.
We adopt \cite{Henon1971} units in the simulations (this is frequently referred to as the $N$-body unit, hereafter we use the abbreviation NB unit).
In these units, the total mass of the cluster ($M$) is unity, the total kinetic energy is $0.25$, and the total potential energy is $-0.5$.
The gravitational constant is $G=1$.
The initial virial radius is one.
The corresponding initial half-mass radius ($R_{\mathrm{h,0}}$) is approximately $0.768$.
The initial crossing time is approximately $0.674$, estimated by using
\begin{equation}
  \tcr = \sqrt{\frac{\rh^{3}}{G M}}.
\end{equation}
The initial relaxation time is approximately 31.038, calculated by using the formula from \cite{Spitzer1987},
\begin{equation}
  \trh \approx 0.138 \frac{N^{1/2} \rh^{3/2}}{m^{1/2} G^{1/2} \ln \Lambda},
  \label{eq:trh}
\end{equation}
where $m=0.001$ and $\Lambda=0.02 N$.
The factor of $0.02$ follows the measurement from \cite{Giersz1996}. %\change{estimate of Lyapunov time from Goodman et al. 1993?}
For the equal-mass Plummer model, core collapse happens after approximately $15\trh$ \citep{Heggie2003}, which corresponds to approximately 466 NB time units for our models.
Thus, we stop the simulations at 500 NB time unit to cover the core collapse process.
When the energy of a particle becomes positive and its distance to the cluster center exceeds 10 NB units, the particle is treated as an escaper.

We carry out 7 groups of models by varying the tree time step ($\dtl$) of the PT method, $\rout$, and $\rg$.
$\dtl$ represents the accuracy of long-range interactions.
$\rout$ and $\rg$ together represent the accuracy of short-range interactions.
Smaller $\rout$ and $\rg$ indicate low accuracy because less particles are treated with the Hermite or SDAR method.

Table~\ref{tab:init} shows the parameters of each group.
For the names of groups, the prefixes, ``L'', ``S'' and ``H'', in the front of ``PT'' and ``PP'' indicate the low, standard, and high accuracy, respectively.
The suffix ``-C'' indicates that the model follows the automatic determination of $\rout$ as \citep[Equation 41 in][]{Wang2020c}:
\begin{equation}
  \rout = 10 \dtl \sigma_{\mathrm{1D}},
  \label{eq:routdts}
\end{equation}
where $\sigma_{\mathrm{1D}}$ is the one-dimensional velocity dispersion of the system.
This relation ensures that $\dtl$ is small enough to resolve the motion of particles inside the changeover region.
Therefore, in the LPT-HPP-C group, the long-range interactions have a low accuracy (a large $\dtl$).  In balance, the short-range interactions have a high accuracy ($\rout$ is about $12.8\%$ the initial virial radius). 

Meanwhile, for the HPP and SPP groups, $\rg=0.8\rin$.
This value ensures that the SDAR method is used for most multiple systems inside $\rin$ while the Hermite integrator mainly handles the interactions inside the changeover region.
This choice can best avoid an abnormal orbital evolution of binaries due to the cumulative error of the Hermite integrator.
In the LPP case, $\rg$ is set to a small value $10^{-8}$ so that most wide multiple systems and encounters are not integrated by the SDAR method.
We cannot set $\rg$ to zero because the formation of a tight binary can result in a very small $\dts$, and thus, the simulation becomes extremely slow.

\begin{table*}
  \centering
  \caption{The accuracy controlling parameters for each group of the $N$-body models. $\dtl$ is tree time step, $\rout$ is the outer boundary of the changeover function. $\rg$ is the radial criterion to switch on the SDAR method. All values are in the NB unit.}
  \label{tab:init}
  \begin{tabular}{cccccccc} 
    \hline
    Groups     & \multicolumn{3}{c}{STD}          & LPT-SPP &   \multicolumn{3}{c}{LPP} \\
    \hline
    Model Name & SPT-SPP-C & LPT-HPP-C & HPT-SPP  & LPT-SPP & LPT-LPP        & SPT-LPP       & HPT-LPP         \\
    \hline                                                                                  
    $\dtl$     & $1/128$   & $1/32$    & $1/1024$ & $1/32$  & $1/32$         & $1/128$       & $1/1024$        \\
    $\rout$    &    0.03   & 0.128     & 0.03     & 0.03    & 0.0001         & 0.0001        & 0.0001          \\
    $\rg$      &    0.0024 & 0.01024   & 0.0024   & 0.0024  & $1\times10^-8$ & $1\times10^-8$& $1\times10^-8$  \\
    \hline
  \end{tabular}
\end{table*}

The 7 groups can be divided into three classes.
The SPT-SPP-C, LPT-HPP-C and HPT-SPP groups have self-consistent (satisfying the criterion determined by Equation~\ref{eq:routdts}) accuracy parameters for long- and short-range interactions. 
We call them as the STD class.
The LPT-LPP, SPT-LPP and HPT-LPP groups have a low accuracy of short-range interactions.
We call them as the LPP class.
The LPT-SPP has the same accuracy parameters for short-range interactions as those in SPT-SPP-C and HPT-SPP groups, but it uses low accuracy parameters for long-range interactions. 
Thus we separate it into the third class.

%\change{Maybe good to note $r_{\mathrm{out}}$ as big as $10\%$ of virial radius in LPT-HPP-C.  By the way, isn't a conclusion from this that long-range forces actually don't really matter?  It seems this should be a highlight of the paper.  We can reference Chris Hamilton's paper and say that we favor the Chandrasekhar relaxation picture, not the alternative one they propose.}
For each group, we carry out $300$ simulations by only varying the values of $\rout$ via adding a 3 digit number.
For example, in the SPT-SPP-C group, the values of $\rout$ for the first, the second and the last simulations are $0.03$, $0.03001$, and $0.03299$, respectively. % \change{Could we have also just perturbed the initial conditions slightly (which is easier)?  Maybe explain why you chose this approach instead.}.
In this way, we ensure that the initial conditions are exactly the same but only the accuracy parameter of the integration is slightly perturbed.
Since the numerical error grows exponentially in a crossing time \citep{Heggie1991}, we can observe a large divergence of the individual orbits of stars soon after the simulations start.
Thus, we can investigate how sensitively the evolution of the star cluster depends on the accuracy parameter.
%In addition, the statistical properties of the evolution measured from all models reflect the \change{PLEASE REWRITE mean trend referring to the base accuracy parameters of the group (Table~\ref{tab:init})}.

\section{Results}
\label{sec:result}

\subsection{Structure evolution}
\label{sec:lagr}

\begin{figure}
  \includegraphics[width=\columnwidth]{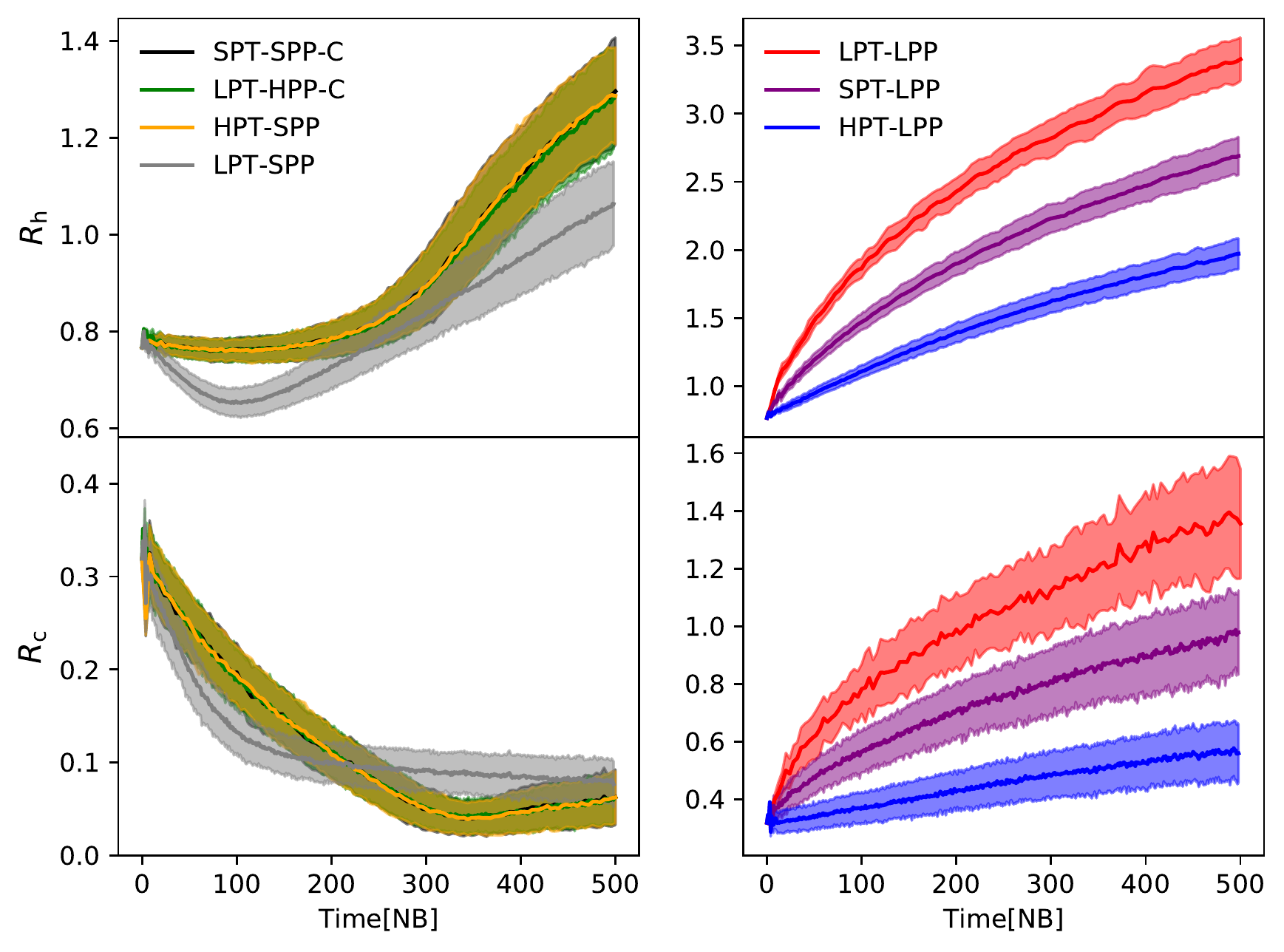}
  \caption{The evolution of the half-mass radius ($\rh$) and the core radius ($\rc$) for all groups.
    The left panels include the STD and the LPT-SPP classes.
    The right panels include the LPP class.
    The central curves are the averaged values of all models in each group.
    The shaded regions show the standard deviations.
  }
  \label{fig:rhrc}
\end{figure}

Figure~\ref{fig:rhrc} shows the evolution of the half-mass radius ($\rh$) and the core radius ($\rc$) for all groups.
The left and right panels compare the STD/LPT-SPP and the LPP classes.
The STD class shows an identical evolution of $\rh$ and $\rc$.
The core collapse finishes around 300 time units.
After that, $\rh$ starts to increase.
The HPT-SPP group has a better accuracy of long-range interactions compared to that of the SPT-SPP-C group; 
this suggests that when $\dtl$ and $\rout$ are set in a self-consistent way, the statistical result can converge and does not depend on the accuracy parameters.

The LPT-SPP group shows a faster core collapse that finishes at approximately 100 time units.
In addition, this group shows a strong decrease of $\rh$ before the end of core collapse.

In contrast, the LPP class shows a completely different evolution of $\rh$ and $\rc$.
The whole system expands immediately after the simulations start, and there is no core collapse.
The expansion indicates that the system is not in virial equilibrium.
Such behaviour may be caused by the artificial energy heating coming from the integration errors of close encounters.
{This indicates that the star cluster has a nonphysical evolution.  For example, in the STD class, before the tight binary, which acts as a heating source, forms, virial equilibrium is the natural state of the star cluster; i.e., $\rh$ does not significantly evolve.
If the cluster is not in virial equilibrium, it will evolve to the virial equilibrium via phase mixing and violent relaxation in a free-fall timescale \citep[e.g.][]{Binney1987}. 
}

This result suggests that without an accurate integration of short-range interactions, the simulation is completely unreliable.
The accuracy of long-range interactions is less crucial but affects the evolution of core and $\rh$.

\subsection{Number of escapers}

\begin{figure}
  \includegraphics[width=\columnwidth]{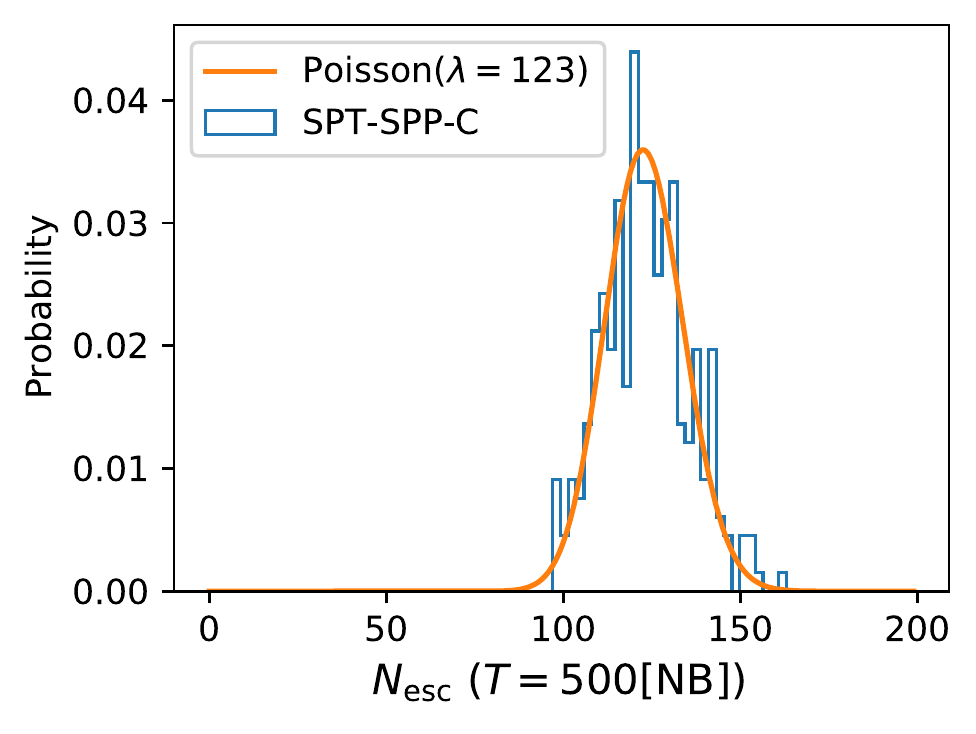}
  \caption{The probability of $\nesc$ for all models in the SPT-SPP-C group.
    A fit of the Poisson distribution is shown in the brown color. The expected value is $\lambda=123$.
  }
  \label{fig:nescstd}
\end{figure}

When the energy of a star becomes positive after encounters, the star becomes a potential escaper.
Then, it travels through the halo to escape, but during travel, distant perturbations can still change its orbit and the star may again become bound.
Thus, the number of escapers ($\nesc$) is sensitive to both long- and short-range interactions.
%\cite{Heggie1991} showed that $\nesc$ has a large difference model by model with different $\eta$ in Equation~\ref{eq:dth}.
In Figure~\ref{fig:nescstd}, we show the probability of $\nesc$ of all models in the SPT-SPP-C group.
%Even without the change of integrating accuracy, 
Similar to the evolution of $\rh$ and $\rc$, a large scatter of $\nesc$ appears among the models.
However, the probability of $\nesc$ approximately follows a Poisson distribution with an expected value of $\lambda=123$.
This suggests that the behaviour of escapers has no correlation with the accuracy parameters; the difference of $\nesc$ is purely due to a random effect.
Therefore, $\nesc$ of individual models is not reliable; to understand the properties of escapers in a low-mass star cluster, it is necessary to obtain ensemble statistics.
\cite{Hernandez2020} found a similar behaviour for the phase space structure of planetary dynamics.

\begin{figure}
  \includegraphics[width=\columnwidth]{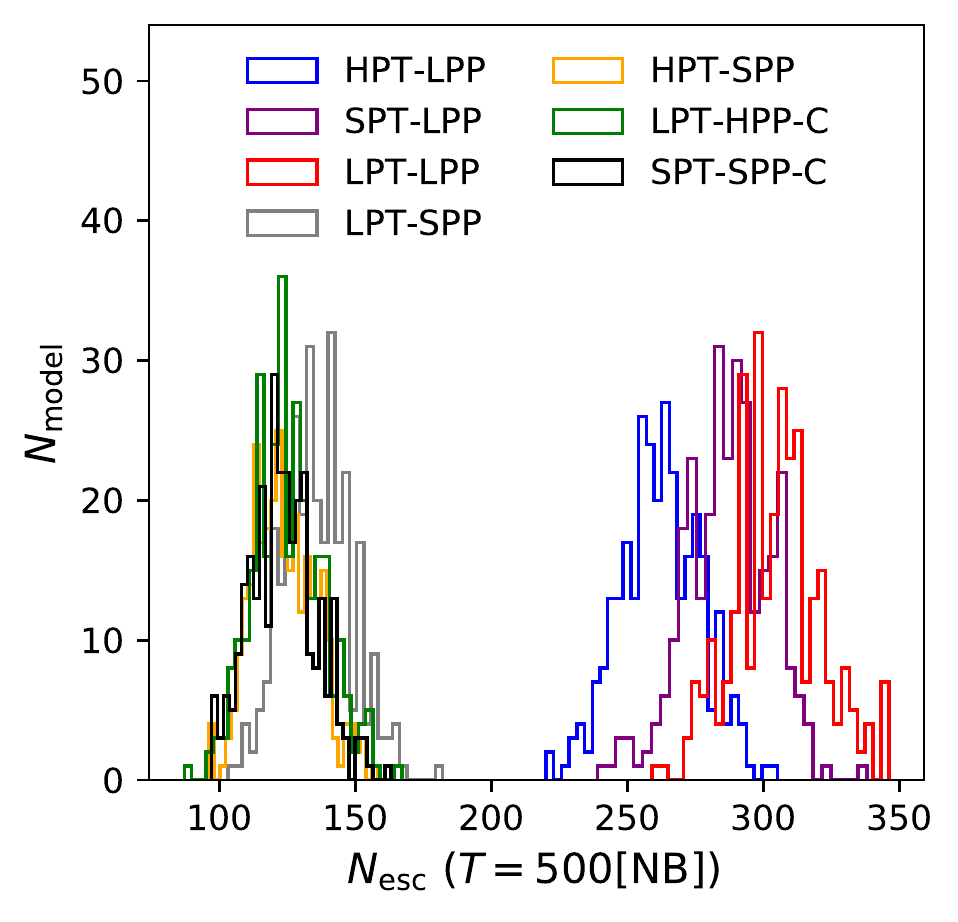}
  \caption{The distribution of $\nesc$ for the models in individual groups.}
  \label{fig:nescgroup}
\end{figure}

Subsequently, we compare the distribution of $\nesc$ from different groups in Figure~\ref{fig:nescgroup}.
The three groups in the STD class well agree with each other.
This is consistent with the behaviour of $\rh$ and $\rc$.
% \change{I think we could explore this quantitatively.  The PDFs can be compared using a 2-sample KS test to see if they are consistent with each other}
%The former two groups use the automatic determination of $\rout$ based on $\dtl$ by using Equation~\ref{eq:routdts}.
The LPT-SPP group has a slightly larger mean $\nesc$.
In contrast, the LPP class with a low-accuracy short-range interaction and different accuracy long-range interactions (different $\dtl$) shows a much larger mean $\nesc$.
This indicates that $\nesc$ is much more sensitive to the accuracy of short-range interactions.
Thus, the accurate treatment of close encounters and multiple systems is crucial for simulating collisional stellar systems.

\subsection{Kinetic energy of escapers}

\begin{figure}
  \includegraphics[width=0.8\columnwidth]{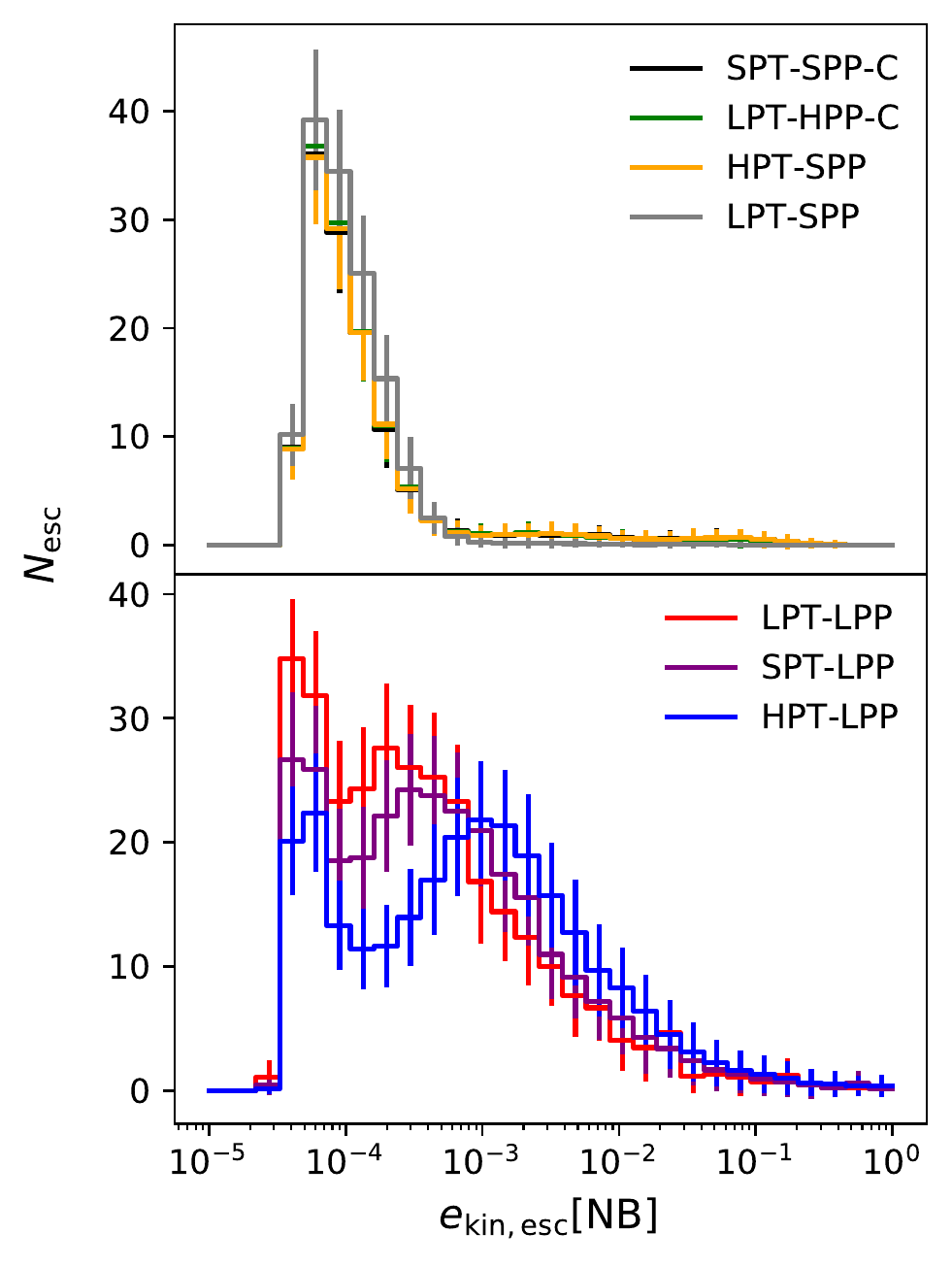}
  \caption{The kinetic energy distribution of escapers.
    The histograms show the averaged value of all models in each group.
    The standard deviations are shown as the error bars.
    The upper panel includes the STD and the LPT-SPP groups. 
    The lower panel includes the LPP class.
    %In order to distinguish different groups, an offset of y-axis is added in each histogram.
    %Thus the value of y-axis does not represent the actual number of escapers.
    %\change{I think there should be a better way to represent these data rather than offsets.  The $Y$-label should be something like PDF, so the axis scale doesn't make much sense currently.}
  }
  \label{fig:ekindist}
\end{figure}

To investigate why the low-accuracy short-range interactions cause a large number of escapers, we plot the distribution for the kinetic energies of escapers in Figure~\ref{fig:ekindist}.
The STD class shows the same distribution again.
Most escapers have a low kinetic energy with a peak around $5\times 10^{-5}$.
There is a long and inconspicuous tail of high-energy ($\eesc>10^{-3}$) escapers.
The LPT-SPP group has an almost identical distribution.
Thus, the low accuracy of long-range interactions does not have an obvious effect on the energy distribution of escapers.

In contrast, the LPP class has a very different distribution with a bimodal shape.
There are more low-energy escapers with $\eesc<10^{-4}$ and a large cluster of high-energy escapers with $10^{-4}<\eesc<10^{-2}$.
The peak of the high-energy part shifts to a higher value when the accuracy of long-range interactions increases.

\begin{figure}
  \includegraphics[width=\columnwidth]{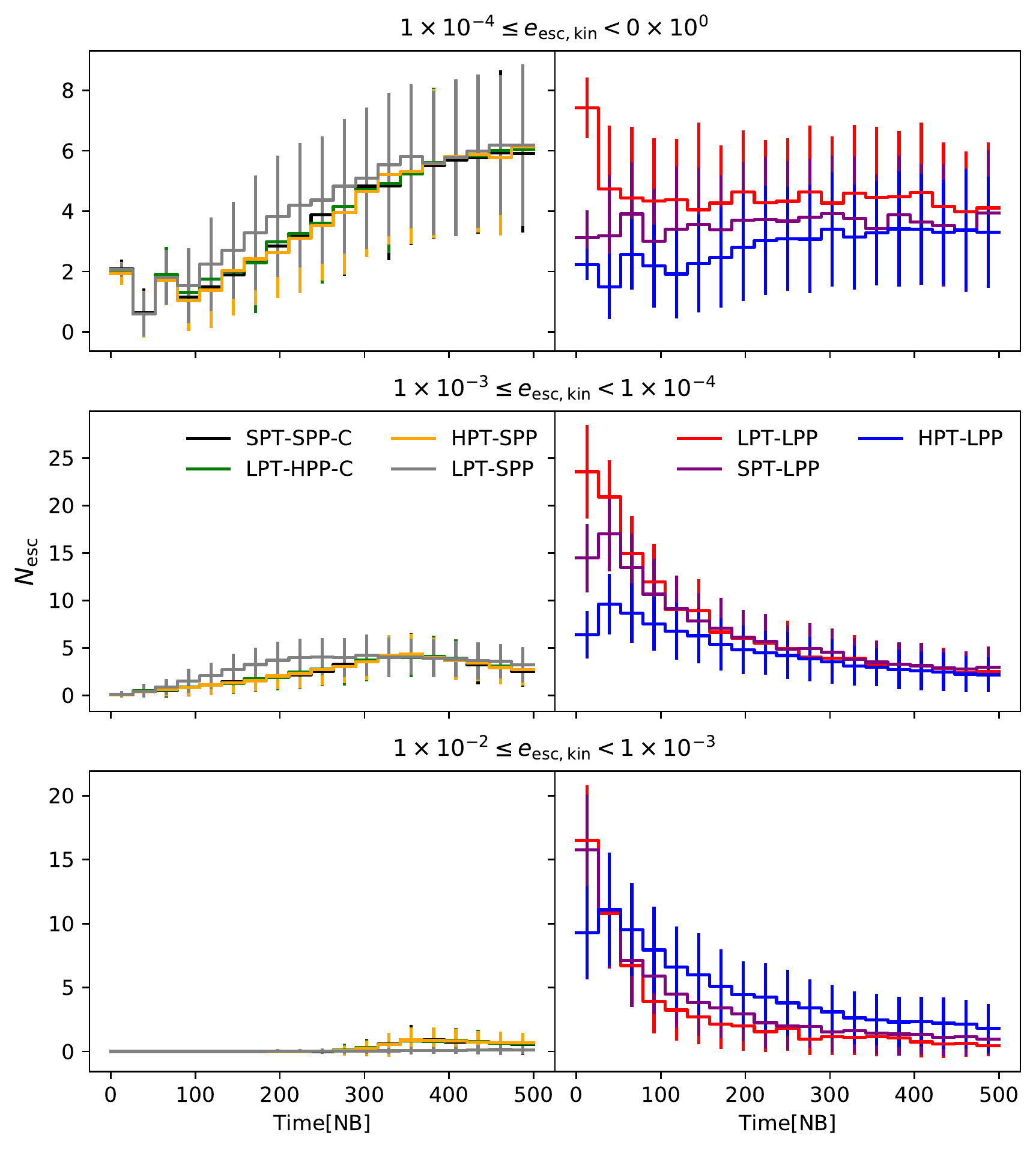}
  \caption{The number of escapers depends on the time: $\nesc(t)$.
    The escapers are collected into to three (low, middle and high) kinetic-energy bins.
    The left panels include the STD and the LPT-SPP classes.
    The right panels include the LPP class.
    The step curves are the mean $\nesc(t)$ of all models in each group.
    The errorbars are the standard deviations.
  }
  \label{fig:nesctime}
\end{figure}

%\change{I think maybe this next figure is not necessary, if one needs to get deleted.}
In order to understand the origin of the bimodal distribution forms, we analyze how $\nesc$ depends on the time as shown in Figure~\ref{fig:nesctime}.
For the STD and the LPT-SPP classes, only a few low-energy escapers appear initially, and then, their number increases.
After 300 time units, the high-energy escapers start to appear.
This is due to core collapse, which leads to the formation of tight binaries.
These binaries can generate high-energy escapers via the decays of unstable multiple systems.
Compared to the STD class, the low- and middle-energy escapers appear slightly earlier in the LPT-SPP group because of an earlier core collapse.

The LPP class (right panel) shows a completely different behaviour.
A large number of escapers appear immediately after the simulations start.
This is the case for all energy bins in the LPT-LPP group, which has the lowest accuracy for both the long-range and the short-range interactions.
These initial escapers are corresponding to the right peaks in the distribution of $\eesc$ (Figure~\ref{fig:ekindist}).
When $\dtl$ decreases, the numbers of low- and middle-energy escapers decrease.
As shown in Figure~\ref{fig:rhrc}, the LPP class has immediately left the virial equilibrium at the beginning. 
The integration errors of close encounters lead to a boost of escapers. 
A higher accuracy of long-range interactions help somewhat.

Meanwhile, if we ignore the initial escapers, the numbers of middle- and high-energy escapers decrease in the later evolution, in contrast to those in the STD class.
There is no core collapse while the density continues to decrease in the LPP class.
Thus, such behaviour is expected. 
The low-energy escapers continue to appear and cause the left peak in the distribution of $\eesc$.

\subsection{Energy error}

The total energy in a Newtonian gravitational $N$-body system is conserved.
Thus, the energy conservation is frequently used to measure the fidelity of $N$-body models.
However, this does not work well for collisional stellar systems.
If a tight binary forms, based on the Heggie-Hills law \citep{Heggie1975,Hills1975}, the maximum semi-major axis can be described as
\begin{equation}
  \ab \approx \frac{G \ma \mb}{\mave \sigma^2},
  \label{eq:ab}
\end{equation}
where $\ma$, $\mb$, and $\mave$ are the masses of the two components and the locally averaged stellar mass, respectively; $\sigma$ is the local velocity dispersion.
In our models, $\ma\mb/\mave=1.0$ and $\sigma\approx 1$ after core collapse for the STD class.
Thus, $\ab\approx 1$ and the corresponding binding energy of the binary is $0.5$.
This is close to the initial total potential energy of the whole cluster.
After many encounters, the binding energy can even become larger.
Then, the total energy of the system is completely dominated by this binary.
The energy error from the binary can mask all other small energy errors in the system.
%Therefore, the energy conservation would not be useful to measure the quality of the integration anymore.

Similarly, when there are still no binaries, the error is dominated by close encounters with a pair separation below $\ab$.
{If a single encounter is not accurately followed, it can create a large energy error that does not affect the dynamical state of the whole system significantly.
However, if every encounter is treated inaccurately, each with small energy error, the error may not add coherently.  The cumulative error may appear small even when the evolution is nonphysical.}

\begin{figure}
  \includegraphics[width=\columnwidth]{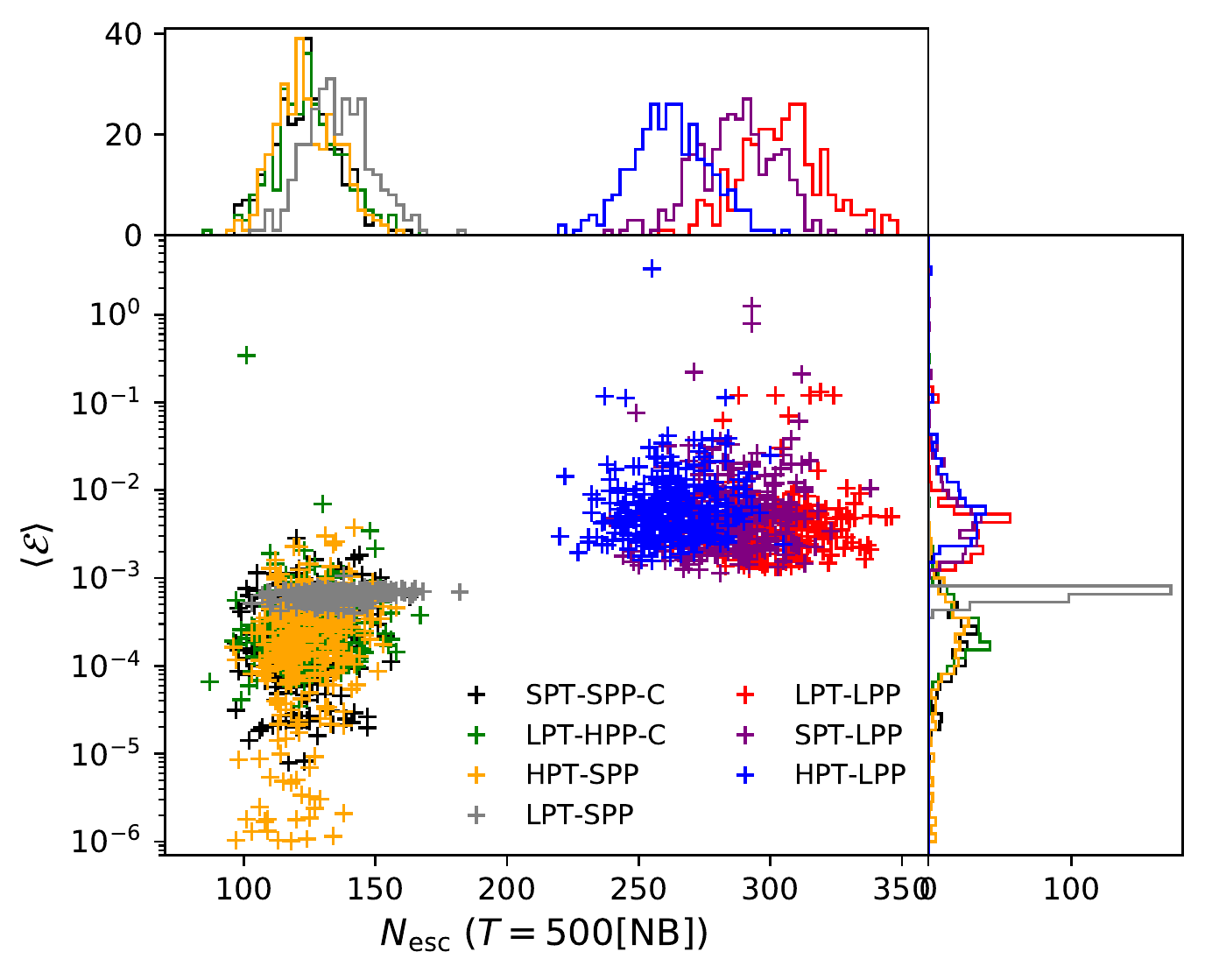}
  \caption{The average energy error per time unit ($\eerr$) vs. $\nesc$ in each model from all groups.
    The upper and right panels show the corresponding histograms.
  }
  \label{fig:nescerr}
\end{figure}

Figure~\ref{fig:nescerr} shows the averaged energy error per time units ($\eerr$) vs. $\nesc$ for all groups.
The three classes show different properties of $\eerr$.
The STD class has a relatively low mean $\eerr$, and some models in the HPT-SPP group have the lowest $\eerr$.
The LPT-SPP group shows a narrow distribution of $\eerr$ with the mean value close to the upper boundary of the STD class.
The LPP class has the largest values of $\eerr$.
For individual models in each group, there is no obvious correlation between $\nesc$ and $\eerr$.
Especially, in the LPP class, the mean $\nesc$ among the groups vary widely, while the distributions of $\eerr$ are similar;
%Meanwhile, the stochastic errors from binaries or close encounters causes a large scatter of $\eerr$. 
some models in the STD class have a large $\eerr$ similar to those in the LPP class.
Therefore, $\eerr$ cannot explain whether a system has a correct physical evolution.

\section{Discussion and conclusion}
\label{sec:conclusion}

In this work, we carry out 7 groups of $N$-body simulations of an equal-mass star cluster with 1000 particles by using the \textsc{petar} code.
We investigate how the accuracy parameters of long-range ($\dtl$) and short-range interactions ($\rout$ and $\rg$) affect the dynamical evolution of the star cluster. 
All simulations use the exact same initial condition.
In each group, we perform 300 integrations with slight modifications to $\rout$ to obtain different starting numerical errors.
We analyze how the structure evolution and the properties of escapers are related to the accuracy parameters of the integrator.

The dynamical evolution, especially the relaxation process, of the clusters are affected by both short-range interactions (close encounters and multiple systems) and long-range interactions.
If the accuracy parameters are set in a self-consistent way (the STD class; see Equation~\ref{eq:routdts}), the statistical properties of the evolution are independent of the accuracy parameters.
When the accuracy is not sufficient for long-range interactions (the LPT-SPP group) compared to that of the STD class, the core collapse occurs earlier and the number of escapers $\nesc$ is slightly larger.
%\change{Hmm, the long-range interaction is on high setting, so I don't understand what you mean.  What I think happens here is that the short-range part does not become engaged self consistently, but this is still a short-range problem; my feeling from the paper is that long-range can be done crudely...}, 
However, when the short-range interaction is inaccurate (the LPP class), the clusters have nonphysical evolution:
the clusters evolve out of virial equilibrium immediately after the simulations start, i.e., there is a strong expansion of $\rh$.
{Since the phase mixing and the violent relaxation drive a stellar system into virial equilibrium, a contrary evolution is not correct.}
In addition, no core collapse occurs and $\nesc$ is significantly larger than that of the STD group.
Therefore, realistic $N$-body simulations of collisional stellar systems must accurately treat short-range interactions.

Although individual models in one group have a large difference of $\rh$, $\rc$ and $\nesc$, the distributions of them in a group converge. For example, the distribution of $\nesc$ follows the Poisson distribution function (Figure~\ref{fig:nescstd}).
Thus, although an initial small difference can grow exponentially so that individual $N$-body models cannot reproduce the exactly same evolution of a real cluster, the statistical means are meaningful.  A similar result in the orbits of planets is found in \cite{Hernandez2020}.

%As the algorithms to deal with the short-range interactions, such as KS regularization in \textsc{nbody6} and the SDAR method in \textsc{petar} sometimes cannot avoid large energy errors of few-body interactions due to the balance requirement of the performance and the accuracy.
%TSince a tight binary is energetic, the integration error from it dominates the total energy error of the system.
The energy error has a large scatter among the models, but the physical properties of the systems do not have a clear dependence on $\eerr$ (Figure~\ref{fig:nescerr}).
This is reasonable because in collisional stellar systems, the integration errors from tight binaries or close encounters dominate the total energy error and dominate over the systematic error generated by the nonphysical evolution.
Thus, checking the energy conservation is not the proper way to measure the quality of $N$-body simulations for collisional stellar systems.
It is much more important to have a proper algorithm to correctly treat the short-range interactions rather than achieving a good energy conservation by using an approximation (e.g., a softening potential).

\cite{Hernandez2020} found that the relevant metric for accuracy was the number of steps per bound orbit, rather than the energy error.  They found that orbits needed at minimum about $16$ steps per effective period at pericenter, a number which depends on eccentricity.  This number holds for a Wisdom--Holman map \citep{WH91}, but nonetheless, we assume this approximate criteria applies to other maps.  Using a maximum of $16$ integration steps per orbit, the minimum semi-major axes of a binary ($a_{\mathrm{min}}$) that the long-range PT integrator can resolve are approximately $0.002$, $0.009$, and $0.02$ NB length units for the HPT, SPT and LPT groups, respectively (see Table~\ref{tab:init}).
In the HPT-LPP group, $\rout=0.0001$, which is $5$ times less than $a_{\mathrm{min}}$.
The situation in the other two groups of the LPP class are even worse.
However, the three groups in the STD class have $\rout>a_{\mathrm{min}}$.
Thus, our result is consistent with what was found in \cite{Hernandez2020}.

%The three groups in the STD class have different combinations of accuracy parameters for the long- and short-range interactions, but the results are identical.
%A low accuracy in the long-range interaction can still reproduce a reasonable result if the accuracy of short-range interactions is determined in a self-consistent way (Equation~\ref{eq:routdts}).

In this work, we only investigate the low-$N$ star clusters, which have a large statistical scatter.
In the case of massive stellar systems such as globular clusters with million stars, the statistical scatter is expected to be much smaller.
Thus, $\nesc$ in one model would be closer to the statistically expected value.

Our analysis is based on the specific algorithms used in \textsc{petar}.
We refer the PT and PP to the long-range and short-range interaction, respectively.
This does not represent all type of $N$-body algorithms.
With different integrators, the results can change, although the general trend is not expected to vary significantly.
%\change{as I mentioned before, I think a conclusion of our paper is that the Chadrasekhar picture of relaxation, which is dominated by close encounters, is supported in this paper, because long-range interactions can be treated crudely.  This maybe should be a highlight in our conclusions.  Let me know if you agree, and we can put a reference to Chris Hamilton's paper.  On the other hand, short-range must always be done correctly}

\section*{Acknowledgements}

L.W. thanks the financial support from JSPS International Research Fellow (School of Science, The university of Tokyo).
Numerical computations were in part carried out on Cray XC50 at Center for Computational Astrophysics, National Astronomical Observatory of Japan.  We thank discussions with Chris Hamilton.

%%%%%%%%%%%%%%%%%%%%%%%%%%%%%%%%%%%%%%%%%%%%%%%%%%
\section*{Data Availability}

The $N$-body simulations underlying this article were generated by using the \textsc{petar} code on the desktop computer of the corresponding author and the supercomputer, Cray XC50 at Center for Computational Astrophysics (CfCA), National Astronomical Observatory of Japan. The data underlying this article will be shared on reasonable request to the corresponding author. 



\bibliographystyle{mnras}

%\begin{thebibliography}{99}
%\end{thebibliography}

%%%%%%%%%%%%%%%%%%%%%%%%%%%%%%%%%%%%%%%%%%%%%%%%%%

%%%%%%%%%%%%%%%%% APPENDICES %%%%%%%%%%%%%%%%%%%%%

%\appendix

% Don't change these lines
\bsp	% typesetting comment
\label{lastpage}
\end{document}